# Molecular Insights into Yb(III) Speciation in Sulfate-Bearing Hydrothermal Fluids from X-ray Absorption Spectra Informed by *ab initio* Molecular Dynamics

Xiaodong Zhao,[1,2] Duo Song,[1] Sebastian Mergelsberg,[1] Micah Prange,[1] Daria Boglaienko,[3] Zihua Zhu,[4] Zheming Wang,[1] Carolyn I. Pearce,[3] Chengjun Sun,[5] Kevin M. Rosso,[1] Xiaofeng Guo,[2] Xin Zhang[1] *

1. Physical & Computational Science Directorate, Pacific Northwest National Laboratory, Richland, Washington 99354, United States
2. Department of Chemistry, Washington State University, Pullman, Washington 99164, United States
3. Energy and Environment Directorate, Pacific Northwest National Laboratory, Richland, Washington 99354, United States
4. Environmental Molecular Sciences Laboratory, Pacific Northwest National Laboratory, Richland, Washington 99354, United States
5. Advanced Photon Source, Argonne National Laboratory, Lemont, Illinois 60439, United States

**Corresponding Authors:** xin.zhang@pnnl.gov (X.Z.)

## Abstract

Rare earth elements (REEs) are critical for advanced technologies, yet in hydrothermal aqueous solutions the molecular level details of their interaction with ligands that control their geochemical transport and deposition remain poorly understood. This study elucidates the coordination behavior of $Yb^{3+}$ in sulfate-rich hydrothermal fluids using *in situ* extended X-ray absorption fine structure (EXAFS) spectroscopy and *ab initio* molecular dynamics (AIMD) simulations. By integrating multi-angle EXAFS with AIMD-derived constraints, we precisely resolve $Yb^{3+}$ coordination structures and ligand interactions under hydrothermal conditions. At room temperature, $Yb^{3+}$ is coordinated by five water molecules and two sulfate ligands (coordination number, CN = 8), forming a distorted square antiprism geometry. Increasing temperature induces progressive dehydration, reducing the hydration shell and favoring stronger sulfate complexation. At 200°C, sulfate ligands reorganize around $Yb^{3+}$, shifting its geometry to a capped dodecahedron (CN = 7). At 300°C, sulfate binding dominates, leading to structural reorganization that parallels the onset of sulfate mineral precipitation, consistent with the retrograde solubility of REE sulfates. These findings provide direct molecular-scale evidence that sulfate acts as both a transport and deposition ligand, critically influencing REE mobility in geochemical environments. Our results can also help to refine thermodynamic models of REE speciation in high-temperature hydrothermal fluids and improve our understanding of REE ore formation processes in nature.

**Key words**: Hydrothermal Fluid, $Yb^{3+}$, sulfate EXAFS, AIMD, REE Sulfate

## 1. Introduction:

The demand for rare earth elements (REEs, a group of 17 elements including Sc, Y, and La to Lu) has surged in recent years due to their critical roles in cutting-edge technologies, including renewable energy, electric mobility, and high-strength alloys. This growing industrial dependency has spurred worldwide efforts to identify and exploit alternative geological sources of REEs.(Atwood, 2013; Balaram, 2019; Chen et al., 2024; Zhao et al., 2024) Hydrothermal fluids play a crucial role in forming economically significant REE ore deposits, as observed in locations such as the Gallinas Mountains in New Mexico (United States), the Snowbird Deposit in Montana (United States), the Pea Ridge in Missouri (United States), and Bayan Obo in Inner Mongolia (China).(Fan et al., 2016; Harlov et al., 2016; Metz et al., 1985; Smith et al., 2000; Williams-Jones et al., 2000) Over the past two decades, experimental and computational studies have advanced our understanding of REE speciation in hydrothermal systems under elevated temperature and pressure. Techniques such as in situ X-ray absorption spectroscopy (XAS) and ab initio molecular dynamics (AIMD) simulations have been instrumental in elucidating the molecular-level coordination environment of REEs, particularly their interaction with ligands such as chloride, sulfate, and fluoride. Early studies, including those by Mayanovic *et al.* (Anderson et al., 2002; Mayanovic et al., 2007, 2009a), demonstrated the efficacy of *in situ* spectroscopic techniques for characterizing REE hydration and ligand exchange processes. Subsequent studies extended this approach to sulfate and fluoride complexes, highlighting how ligand type, temperature, and pressure affect REE mobility (Guan et al., 2020; Migdisov and Williams-Jones, 2008).

Ligands such as $Cl^-$, $CO_3^{2-}$ and $SO_4^{2-}$ have been considered mainly as transporting ligands in nature, while $F^-$ and $PO_4^{3-}$ have been considered as depositional ligands. (Louvel et al., 2022; Migdisov et al., 2016) Among these, $SO_4^{2-}$ was considered as dominant transporting ligand in certain REE deposits, such as Bayan Obo (China) (Lai and Yang, 2013), Maoniuping (China) (Zheng and Liu, 2019), Mountain Pass (United States) depositions (Verplanck et al., 2016). In hydrothermal fluid, the solubility of REE sulfates undergoes a unique retrograde dissolving behavior; the solubility decreases with increasing temperature, enabling precipitation at elevated temperatures. However, in the presence of prograde soluble alkali sulfates (e.g., $Li_2SO_4$ and $Na_2SO_4$) (>250°C and 90 MPa) the increased alkali sulfate solubility may enhance the stability of REE sulfate complexes, especially $REE(SO_4)^{2-}$ and $REESO_4^+$.(Wan et al., 2023) Furthermore, the recent discovery of liquid–liquid immiscibility (LLI) in REE sulfates revealed the potential for $REE^{3+}$-$SO_4$ and $REE^{3+}$-$HSO_4$ complexes to become concentrated into a dense liquid phase (DLP) and transport effectively.(Wan et al., 2021) These findings underscore the significant role of sulfate in transporting and enriching REEs in hydrothermal environments.

Despite recent advances, the molecular-level structural dynamics of REE sulfates under high-pressure and high-temperature (P−T) conditions remains poorly understood.(Migdisov et al., 2019;

Migdisov et al., 2016; Migdisov and Williams-Jones, 2008) A molecular-scale understanding is crucial for understanding and properly interpreting speciation and consequently transport and deposition mechanisms in various hydrothermal environments. Furthermore, high P−T structural data for REE complexes in aqueous solutions contribute to the development of robust thermodynamic databases that are essential for refining predictive geochemical models.

The primary reactions governing REE sulfate complexation in solution are as follows:(Migdisov and Williams-Jones, 2008)

$$REE^{3+} + SO_4^{2-} \rightleftharpoons REESO_4^+ \qquad (1)$$

$$REE^{3+} + 2SO_4^{2-} \rightleftharpoons REE(SO_4)_2^- \qquad (2)$$

$$REESO_4^+ + SO_4^{2-} \rightleftharpoons REE(SO_4)^{2-} \qquad (3)$$

However, molecular-scale studies of heavy REE sulfate complexes remain scarce, and to the best of our knowledge no structural data exist for the Ytterbium (Yb)-SO₄ system. Key questions remain unanswered, such as whether Yb³⁺ preferentially binds mono-sulfate or bi-sulfate species, and how parameters such as bond angles, bond distances, and binding modes vary under high P−T conditions.

To investigate the molecular structure of Yb-SO₄ complexes under elevated P−T conditions, we employed extended X-ray absorption fine structure (EXAFS) spectroscopy coupled with a compatible hydrothermal diamond anvil cell (HDAC). However, the integration of DACs with energy-scanning spectroscopic techniques has long presented challenges, including distortions in XAS spectra caused by Bragg reflections, limited sample thickness, and significant x-ray absorption by the diamond at lower energies.(Ohsumi et al., 1986; Sueno et al., 1986) In particular, Bragg glitches induced by the DAC often exceed the amplitude of EXAFS oscillations, significantly degrading the quality of the data required for accurate structural analysis. (Ingalls et al., 1980; Ohsumi et al., 1986; Sueno et al., 1986) Several approaches have been proposed to mitigate these artifacts, including measuring multiple spectra at different cell orientations,(Freund et al., 1989; Ingalls et al., 1980; Liu et al., 2020) drilling holes in the diamond for fluorescence mode measurements,(Mayanovic et al., 2007, 2009b; Mayanovic et al., 2002, 2003) using polycrystalline boron carbide ($B_4C$) anvils,(Freund et al., 1989; Sueno et al., 1986) employing low atomic number gasket,(Hu et al., 1994; Kaindl et al., 1988) and using large-volume multianvil apparatus.(Kaindl et al., 1988; Katayama et al., 1997; Liu et al., 2020; Louvel et al., 2022) However, many of these approaches require specialized equipment or involve hazardous materials (e.g., beryllium gaskets). In this study, we adopted the simple multi-angle approach to eliminate Bragg glitches. By rotating the liquid cell up to ±4° relative to the incident beam direction within the plane of synchrotron radiation, we shifted the Bragg peaks in energy space, allowing glitch-containing regions in one scan to be replaced with glitch-free data from another.

To investigate the coordination environment of the absorbing atom, the shell-by-shell fitting method was used for EXAFS analysis. This iterative approach involves adding coordination shells characterized by coordination number (CN), absorber–back scatterer distance (R), and Debye–Waller factor ($\sigma^2$), all of which contribute to the normalized EXAFS spectrum as shown in Equation (4):

$$\chi(k) = \sum_R \frac{S_0^2 CN}{kR^2} |f(k)| e^{-\frac{2R}{\lambda(k)}} e^{-2\sigma^2 k^2} \times \sin(2kR + \phi(k)) \qquad (4)$$

This method is inherently prone to uncertainties and non-unique solutions by the high number of degrees of freedom. Therefore, we improved the accuracy and reliability of the fitting by simulating EXAFS spectra using AIMD-derived trajectories and comparing them with experimental results. This approach of introducing physically meaningful constraints to the fitting enables a more robust interpretation of the data. (Kerisit and Prange, 2019; Xu et al., 2024).

By integrating in situ EXAFS measurements with AIMD simulations, this study provides first molecular-level insights into the structural evolution of $Yb^{3+}$ sulfate complexes under hydrothermal conditions. The results reveal thermally induced dehydration, structural reorganization, and systematic changes in how strongly sulfate binds to $Yb^{3+}$ with temperature, adding mechanistic insight into the role of sulfate in REE transport and deposition in geological environments.

## 2. Material and methods:

### 2.1. The hydrothermal diamond anvil cell

The HDAC is a versatile instrument designed for high P–T studies of hydrothermal solutions (Bassett et al., 1994). It features two opposing 1/8-carat diamond anvils mounted on tungsten carbide seats, which are affixed to upper and lower stainless-steel platens. These platens are drawn together by three screws and precisely aligned by three guide rods, ensuring stability and reproducibility under extreme P–T conditions. Due to the small sample size and the exceptional thermal conductivity of diamond, temperature gradients within the HDAC are minimal in both vertical and horizontal orientations. As a result, temperature corrections during experiments are typically within ±5 °C. The HDAC is mounted on a rotation stage, allowing for rotation about two orthogonal axes perpendicular to the incident X-ray beam to optimize experimental geometry. The sample, comprising a fluid phase (solution and vapor bubble), is confined within a chamber created by a 500 μm diameter hole in a 50 μm thick rhenium gasket. The diamond anvils, with 1 mm diameter flat faces, compress the gasket to seal the sample chamber and maintain the desired pressure.

## 2.2. Sample preparation

Ytterbium(III) sulfate octahydrate ($Yb_2(SO_4)_3 \cdot 8H_2O$, 99.9% purity) was purchased from Thermo Fisher Scientific Inc. (United States). A 0.05 M solution was prepared on-site at the Advanced Photon Source (APS) synchrotron facility by dissolving $Yb_2(SO_4)_3 \cdot 8H_2O$ in deionized water. The solution samples were then loaded into the HDAC chamber using a micropipette under a binocular microscope. While some overflow of the sample solution was often unavoidable, it did not affect the X-ray measurements.

## 2.3. XAS data acquisition and processing

XAS data at the Yb $L_3$-edge (8944 eV) was collected on the 20-BM-B beamline at APS, Argonne National Laboratory (**Figure 1a**). The measurements were performed in transmission mode, with the incident and transmitted X-ray intensities detected using $N_2$-gas-filled ionization chambers. To minimize stray radiation around the focused X-ray beam, a pinhole aperture was positioned in front of the incident-beam ionization chamber. At each P–T set, up to eight scans were acquired, with each spectrum requiring approximately 30 minutes. As shown in **Figure 1**, in instances where Bragg diffraction peaks were observed, the sample cell was rotated by up to ±4° to shift the diffraction peaks in energy space. This adjustment facilitated the replacement of spectral regions containing Bragg peaks with corresponding regions devoid of peaks, ensuring high-quality data for analysis. After Bragg peak removal, the corrected spectra obtained at different detector angles were merged into a single high-quality dataset. This merging process averaged out angle-dependent variations and reduced noise, resulting in a robust and artifact-free spectrum suitable for *k*-space and *R*-space analysis. To further validate the processed synchrotron data, comparative XAFS measurements were performed using a benchtop X-ray spectrometer. The agreement between the benchtop and synchrotron datasets confirms the reliability of the processed synchrotron data, enabling its use for detailed structural analysis of the Yb-SO$_4$ system under high-pressure and high-temperature conditions.

Bench-top EXAFS measurements were collected at room temperature (RT) as a sanity check using an easyXAFS300 instrument (easyXAFS, WA) located at Pacific Northwest National Laboratory (PNNL). The spectra were obtained with a Si(422) spherically bent crystal analyzer and a Mo anode X-ray tube operating at 500 W. Data were corrected for detector deadtime, and the energy was calibrated using a Cr foil standard. Additional spectra were collected with a Ge(422) spherically bent crystal analyzer and a W anode X-ray tube operating at 100 W. Details about these instruments can be found in Seidler *et al.* (Seidler et al., 2014) The data processing and modeling was performed in the Athena and Artemis programs (version 0.9.26).(Ravel and Newville, 2005) The spectra were collected using a Si(551) spherically-bent crystal analyzer and Mo anode x-ray tube operating at 30 kV voltage and 25 mA current. The silicon drift detector deadtime was kept below 30%. The energy calibration of the instrument was done with Cu K-edge (8979 eV). A spectrum collected on a copper foil standard (procured from Exafs materials, CA) was aligned to the one collected at NSLS X11A and retrieved from the Hephaestus database (Ravel and Newville, 2005). Theta-to-energy correction was applied to all the spectra, using theta-to-energy nonlinear shift of the copper standard foil. All the spectra were deadtime corrected. XAS data collection was carried out using an alternating manner between the sample (*It*) and blank (*I0*) to account for possible effects of minor beam instabilities with time. A liquid cell, custom made for the easyXAFS300 instrument, was used for the 0.05 M Yb L$_3$-edge XANES and EXAFS data collection. A small amount of the sample was injected using a syringe attached to one of the PEEK capillary tubes via a PEEK Lure-Lok adapter while the gas head-space was being released through the other tube. The liquid cell had an aperture of 1 cm, flat and parallel x-ray windows (polyimide or Kapton windows) for uniform pathlength that were sealed by a compression mechanism. Absorption length was controlled by a spacer with a thickness of 1.5 mm. The data processing and modeling was performed in the Athena and Artemis programs (version 0.9.26) (Ravel and Newville, 2005).

## 2.4. Computational methods

Density functional theory (DFT)(Becke, 2014) and ab initio molecular dynamics (AIMD) simulations were performed with the pseudopotential plane-wave NWPW module (Bylaska et al., 2011) implemented in the NWChem software package.(Apra et al., 2020; Valiev et al., 2010) The Perdew–Burke–Ernzerhof (PBE)(Perdew et al., 1996) functional was used to account for the exchange correlation energy. Long-range dispersion forces were included in our calculations using the Grimme DFT-D3 method.(Grimme et al., 2011) In our plane-wave calculations, the valence electron interactions with the atomic core were approximated with generalized norm-conserving Hamann pseudopotentials (Hamann, 1989) for H, O, and S. These pseudopotentials were constructed using the following core radii: $r_{cs} = r_{cp} = 0.8$ a.u. for H; $r_{cs} = r_{cp} = r_{cd} = 0.7$ a.u. for O; and $r_{cs} = 0.843$ a.u., $r_{cp} = r_{cd} = 0.960$ a.u. for S. Norm-conserving Troullier-Martins pseudopotentials,(Troullier and Martins, 1991) which contain 4f, 5s, 5p, 5d, 6s and 6p projectors, were applied for Yb. The following core radii were used to generate these pseudopotentials: $r_{cs} = 1.618$ a.u., $r_{cp} = 1.828$ a.u., $r_{cd} = 1.213$ a.u., and $r_{cf} = 1.403$ a.u.. All the pseudopotentials were modified to the separable form suggested by Kleinman and Bylander.(Kleinman and Bylander, 1982) Unrestricted calculations were performed since all the systems were open-shell. The electronic wavefunctions were expanded using a plane-wave basis with periodic boundary conditions at the Γ-point with a wavefunction cutoff energy of 544 eV and a density cutoff energy of 1088 eV.

In AIMD, the system was propagated in time using the Car-Parrinello molecular dynamics (CPMD) scheme.(Car and Parrinello, 1985) The electronic density is described by Kohn-Sham orbitals within the framework of density functional theory (DFT), and these orbitals are linear combination of a plane-wave basis set. The atomic motion during the molecular dynamics is represented by fictious mass. In this work, the simulations were conducted using the canonical ensemble (NVT), with the density of the solution chosen according to the equation of state for NaCl(aq) at similar chloride concentrations and the target temperature and pressure (Driesner, 2007; Driesner and Heinrich, 2007). Equation of motions in CPMD were integrated using position Verlet algorithm,(Verlet, 1967) with a time step 0.12 fs and fictitious orbital mass 600.0 au. All hydrogen atoms were replaced by deuterium to facilitate integration. Temperature control was achieved using the Nosé-Hoover chain thermostat (Blöchl and Parrinello, 1992; Hoover, 1985; Nosé, 1984) for both ions (300 K, 373 K, and 473 K) and electrons (1200 K), and periodic boundary conditions were applied to eliminate surface effects. Initial box was created by Packmol.(Martínez et al., 2009) Each MD simulation was run for 16–30 picoseconds (ps) to ensure sufficient statistical accuracy. The time-averaged stoichiometric and geometric data were obtained using the Visual Molecular Dynamics (VMD) software.(Humphrey et al., 1996) Coordination numbers were derived from the radial distribution functions (RDF), with errors estimated from the differences in RDF integral values at distances of 2.75 Å and 3.25 Å for Yb-O and 2.9 Å and 3.2 Å for Yb-S. The energy and temperature convergency of AIMD results can be found at **Figure S1 - S3**.

## 3. Results

### 3.1. Validation of synchrotron EXAFS data

During EXAFS data processing, we employed a spectral correction approach to remove Bragg peaks introduced by the diamond anvils during synchrotron experiments (as described in the methodology section and shown in **Figure 1**). These Bragg peaks could potentially distort the absorption spectra that are critical for EXAFS analysis. To validate our processed synchrotron data, we conducted an RT experiment using the same sample in a liquid cell with a benchtop easyXAFS300 instrument (easyXAFS, WA). Although this liquid cell cannot sustain the high pressures achievable with HDACs, it effectively eliminates Bragg peaks associated with diamond anvils. A comparison of the EXAFS spectra obtained from the synchrotron and the benchtop instrument in k-space (**Figure 1C**) reveals consistent oscillation patterns across the k-range of 2 to 7 Å$^{-1}$, with minimal deviations in amplitude and phase. This consistency indicates that no significant artifacts were introduced during the Bragg peak removal process. Slight variations in intensity at higher *k*-values are likely due to differences in instrumental resolution or noise levels, as the synchrotron provides higher energy resolution compared to the benchtop instrument. Nevertheless, the overall agreement underscores the robustness of the data validation process and confirms that the synchrotron EXAFS spectra can be reliably used for studying the Yb-SO$_4$ system under high-pressure and high-temperature conditions.

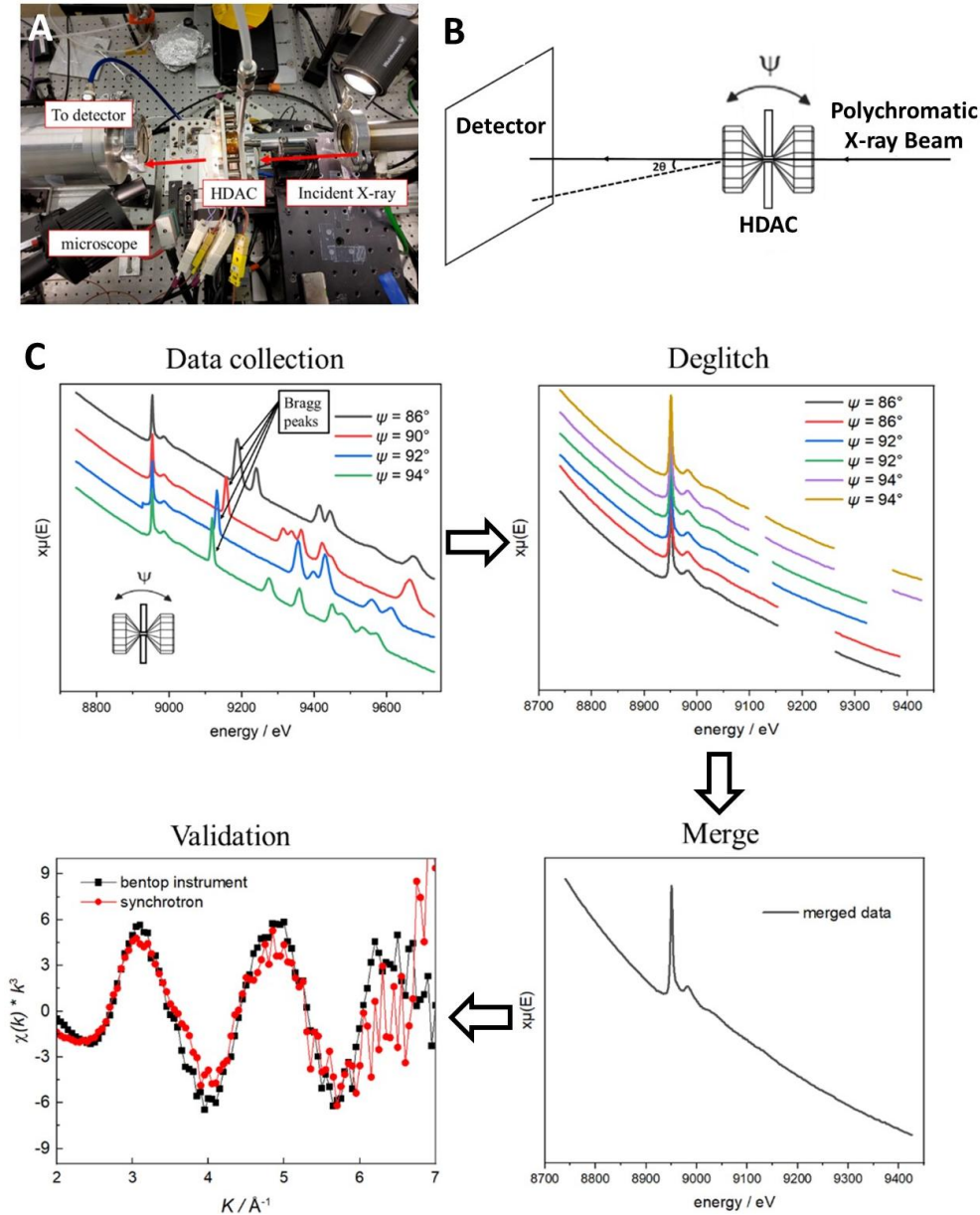

**Figure 1.** (A) Experimental setup for EXAFS measurements at APS. The HDAC is mounted on a rotation stage. The incident X-ray beam passes through the HDAC and the sample, and the transmitted beam is directed to the detector. A microscope is used for alignment and monitoring the sample within the HDAC. (B) Schematic of the HDAC and its rotation during EXAFS measurements. The detector angle ($\psi$) varies to minimize artifacts caused by Bragg peaks from the diamond anvils. (C): Workflow for EXAFS data processing and validation. First step: Raw EXAFS spectra collected at various detector angles at RT ($\psi$=86° to 94°) showing Bragg peaks caused by diamond anvils. Second step: Corrected spectra after Bragg peak removal, demonstrating clean absorption profiles across the energy range. Third step: Merged EXAFS spectra combining data from multiple angles to reduce noise and ensure spectral consistency. Fourth step: Validation of synchrotron data by comparison with benchtop EXAFS spectra in $k$-space, showing consistent oscillation patterns across the $k$-range (2–7 Å$^{-1}$).

## 3.2. Local atomic structure and coordination dynamics of $Yb^{3+}$ from experiments

The experimental results from RT to 200°C, fitting and simulated spectra are shown in **Figure 2** and **Figure 3**, and the fitted results are summarized in **Table 1**. At RT, the experimental $k$-space data agreed well with the simulated spectrum, with the geometry from the simulation was used as the initial model for the shell-by-shell fitting of EXAFS. The $R$-space analysis shows distinct peaks, with the first-shell Yb-O peak appearing at ~2.33 Å, corresponding to a coordination number of ~8.4–8.5. Indicated by the AIMD simulation, this shell consists of five water molecules and three sulfate oxygen atoms, as one sulfate ligand exhibited a bidentate mode. The sharpness of the peaks indicates minimal thermal disorder, supported by the low $\sigma^2$ value of 0.006 Å² for the Yb-O shell. The Yb-S peak at ~3.64 Å corresponds to a coordination number of ~2.5, suggesting two sulfate ligand binding with $Yb^{3+}$. The $\sigma^2$ value for Yb-S at RT is also relatively low (0.006–0.014 Å²), highlighting the rigidity of the sulfate coordination shell at ambient conditions.

**Table 1.** Curvefit Parameters for Yb L-III edge EXAFS from RT to 200°C

| Temperature | Path | CN # | R/Å | $\sigma^2$/ Å² | Temperature | Path | CN # | R/Å | $\sigma^2$/ Å² |
|---|---|---|---|---|---|---|---|---|---|
| RT (bench) | Yb-O | 8.4 | 2.33(3) | 0.006(1) | 100°C | Yb-O | 8.5 | 2.35(1) | 0.010(3) |
| RT (bench) | Yb-S | 2.5 | 3.66(1) | 0.006(8) | 100°C | Yb-S | 2.5 | 3.68(16) | 0.017(31) |
| RT (synchrotron) | Yb-O | 8.5 | 2.32(1) | 0.010(1) | 200°C | Yb-O | 7.8 | 2.34(2) | 0.017(3) |
| RT (synchrotron) | Yb-S | 2.6 | 3.64(2) | 0.014(9) | 200°C | Yb-S | 1.7 | 3.64(1) | 0.006(20) |

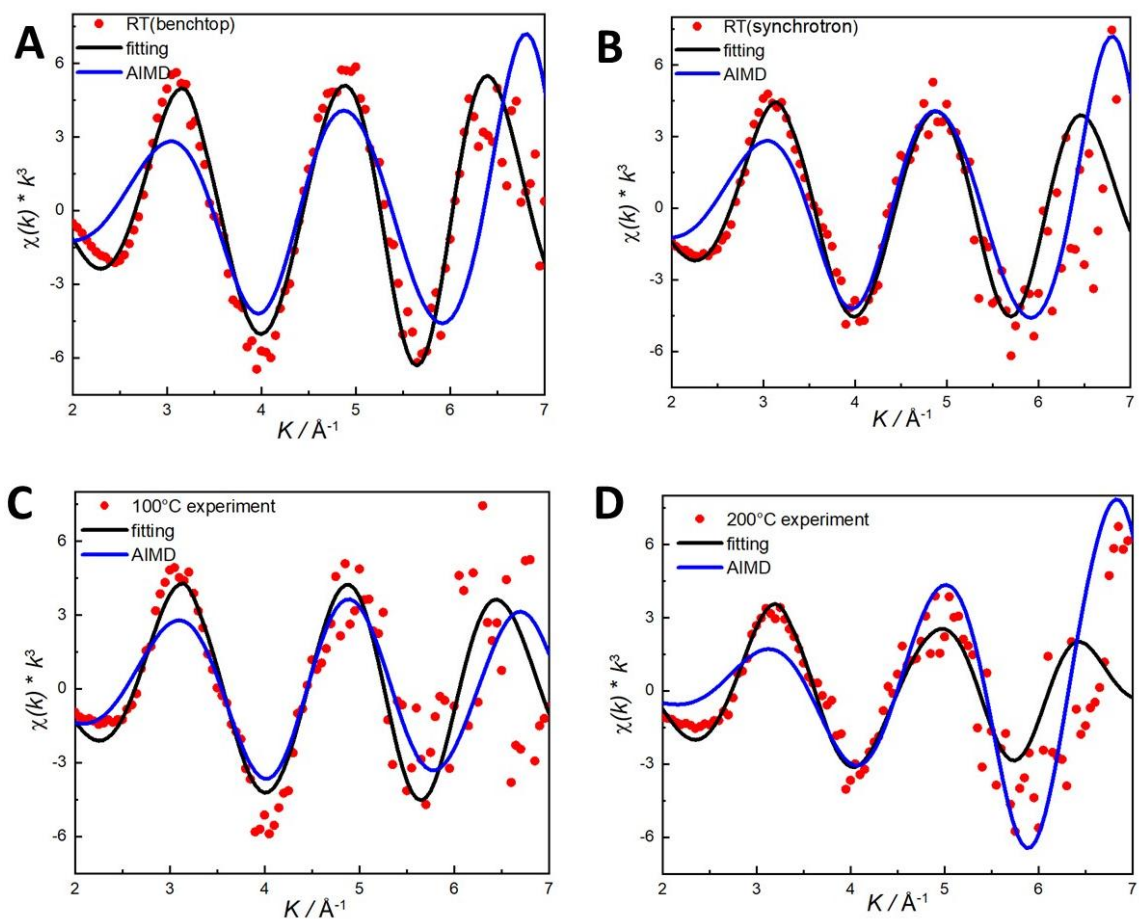

**Figure 2.** Yb L-III edge EXAFS for 0.05M Yb$_2$(SO$_4$)$_3$, shown in $k^3$-weighted k-space from 2 to 7 Å$^{-1}$. (A) RT data measured with a benchtop instrument (red points), theoretical curve fit (black line), and AIMD simulation (blue line). (B) RT data measured at a synchrotron (red points), theoretical curve fit (black line), and AIMD simulation (blue line). (C) Data collected at 100°C (red points), theoretical curve fit (black line), and AIMD simulation (blue line). (D) Data collected at 200°C (red points), theoretical curve fit (black line), and AIMD simulation (blue line). The consistency across experimental and simulated data highlights the accuracy of the modeling and validation processes.

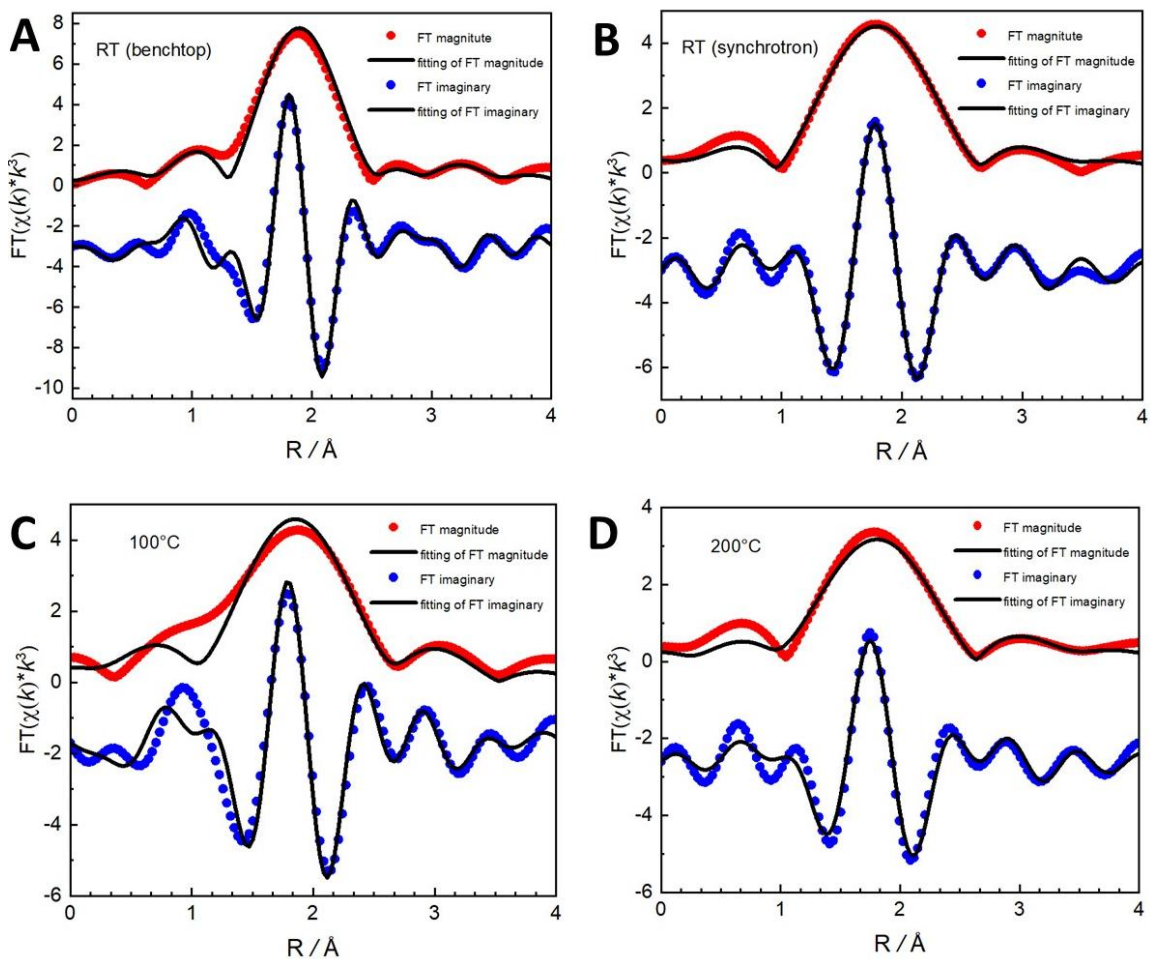

**Figure 3**. Fourier-transformed Yb L-III edge EXAFS for 0.05M $Yb_2(SO_4)_3$ in *R*-space and not phase-corrected. (A) Data collected at RT using a benchtop instrument: red points represent the Fourier-transformed magnitude, blue points the Fourier-transformed imaginary component, and black lines the curve fits for both. (B) Data collected at RT using a synchrotron instrument with the same labeling scheme. (C) Data collected at 100°C with matching Fourier-transformed components and curve fits. (D) Data collected at 200°C matching Fourier-transformed components and curve fits.

As the temperature increases to 100°C, the first-shell Yb-O peak in *R*-space shifts slightly to ~2.35 Å accompanied by peak broadening and a slight increase in the Debye-Waller factor to 0.010 Å². These changes suggest an elongation of Yb–O bonds and increased structural disorder. Despite these changes, the CN remains stable at ~8.5, with $Yb^{3+}$ retaining its hydration shell of five water molecules and sulfate ligands. For Yb-S, the peak shifts slightly to ~3.68 Å, with a CN of ~2.5 and a σ² value of 0.017 Å². This indicates that sulfate ligands remain bound but exhibit slightly more flexibility compared to RT. At 200°C, significant structural changes are observed, driven by dehydration and reorganization of the coordination shell. The $\chi(k)*k^3$ oscillations are noticeably broader, reflecting greater thermal disorder. In the *R*-space, the first-shell Yb-O peak shifts further to ~2.34 Å with diminished intensity, indicating a reduction in the number of coordinating oxygen atoms. The CN decreases to ~7.8 with one water molecule removed from the hydration shell suggested by the AIMD simulation. The $\sigma^2$ value for the Yb-O shell increases to 0.017 Å$^2$, implying greater thermal motion and bond variability. Meanwhile, the Yb-S peak becomes less intense, with the CN decreasing to ~1.7 and the $\sigma^2$ value dropping to 0.006 Å$^2$. This suggests that sulfate groups are still coordinating with Yb by a rigid bond, which is potentially due to stronger interactions with the reduced number of sulfate ligands.

Overall, the structural evolution of Yb$^{3+}$ from RT to 200°C reflects the interplay between hydration, sulfate binding, and thermal effects. At RT, the Yb coordination environment is characterized by a dense hydration shell and stable sulfate binding. As the temperature increases to 100°C, the coordination shell remains intact, with only minor elongation of Yb-O bonds and slight increases in disorder. However, at 200°C, the Yb-S bond length (~3.64 Å) remains stable, and the decreased $\sigma^2$ value indicates that the remaining sulfate ligands bind more rigidly to Yb$^{3+}$. The loss of coordinating water molecules reduces competition for coordination sites and enhances electrostatic interactions between Yb$^{3+}$ and the divalent sulfate. Additionally, the bi-dentate binding mode in sulfate can effectively chelate the Yb with a stable and rigid framework. The structural reorganization at high temperatures further favors sulfate ligands occupying optimal positions despite the reduced coordination number. These observations are consistent with the picture that sulfate serves as a transport ligand at high temperature, at least until 200°C.

At 300°C, the EXAFS indicates a distinct precipitation condition for the Yb-SO$_4$ system. The *k*-space data (**Figure 4A, Table 2**) shows dampened oscillations compared to lower temperatures, indicating greater thermal disorder and possible structural reorganization. The *R*-space data (**Figure 4B, Table 2**) demonstrates broadened peaks, reflecting increased variability in bond lengths and reduced coordination numbers. More importantly, the total CN drops to ~7.2 and the Yb-S CN slightly decreases to ~3.5, which suggests a great dehydration of the complex while yet retaining sulfate ligands. These changes suggest that sulfate ligands bind more strongly with Yb$^{3+}$ as the hydration shell diminishes, likely driving the formation of more ordered sulfate-rich solids, which is consistent with the retrograde solubility behavior of Yb-SO$_4$.(Cetiner et al., 2005; Cui et al., 2020; Wan et al., 2023) Due to the resolution limitation of the optical scope in the experimental set-up, the exact precipitation temperature could not be determined. Previous solubility data did not cover temperature above 200°C,(Judge et al., 2023). Hence, we could only conclude that the precipitation temperature is between 200°C and 300°C, further experiments are highly recommended to resolve a precise precipitation temperature. Comparisons to diffraction data for crystalline Yb$_2$(SO$_4$)$_3$·8H$_2$O indicate that the high-temperature coordination environment approaches that of a crystalline solid, where Yb-S bonds are dominant (CN ~4, bond length ~3.60 Å). The reduced CN for Yb-O at 300°C and the dominance of Yb-S coordination suggests the formation of a partially dehydrated crystal.

**Table 2.** Curvefit Parameters for Yb L-III edge EXAFS at 300°C and bond lengths from diffraction data of $Yb_2(SO_4)_3 \cdot 8H_2O$ solid at RT.

| system | Path | CN # | R/Å | $\sigma^2$/ Å$^2$ |
|---|---|---|---|---|
| HDAC (300°C) | Yb-O | 7.2 | 2.33(8) | 0.012(1) |
| HDA (300°C) | Yb-S | 3.5 | 3.50(12) | 0.037(19) |
| $Yb_2(SO_4)_3 \cdot 8H_2O$ (RT)* | Yb-O | 8 | 2.24 – 2.44 | N/A |
| $Yb_2(SO_4)_3 \cdot 8H_2O$ (RT)* | Yb-S | 4 | 3.60 | N/A |

* Diffraction data of crystal solid from X-ray diffraction by L Hiltunen et al. (Hiltunen and Niinisto, 1976)

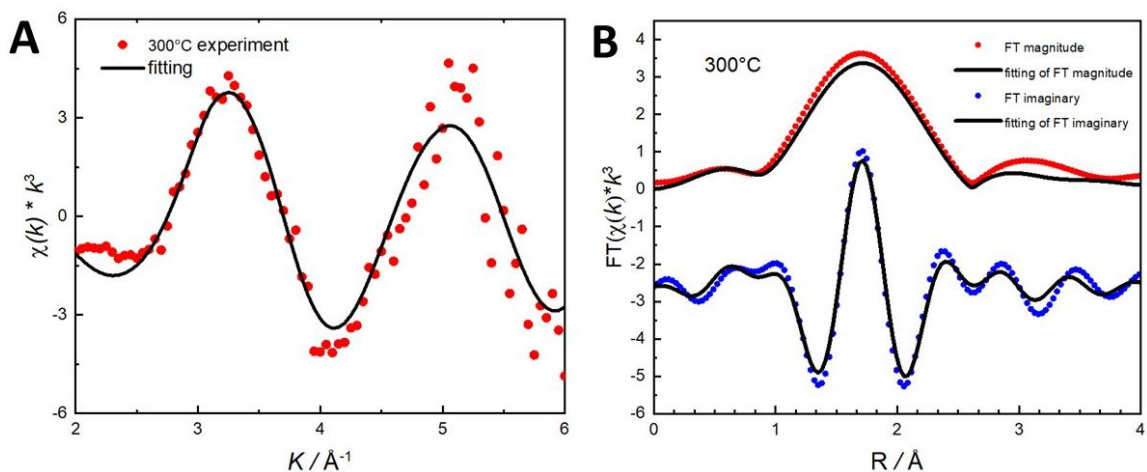

**Figure 4.** (A) Yb L-III edge EXAFS (red points) and the curvefit (line) for 0.05M $Yb_2(SO_4)_3$ at 300 °C, shown in $k^3$ weighted k-space. (B) Yb L-III edge EXAFS (red points for the Fourier-transformed magnitude, and blue points for the Fourier-transformed imaginary components) and the curvefit (line) for 0.05M $Yb_2(SO_4)_3$ at 300 °C, shown in $k^3$ weighted $R$-space and not phase-corrected.

### 3.3. Local atomic structure and coordination dynamics of $Yb^{3+}$ from AIMD

At RT, the coordination environment of $Yb^{3+}$ is defined by a dense hydration shell and strong sulfate binding. The radial distribution function (RDF) for Yb–O (**Figure 5A**) shows distinct peaks corresponding to five water molecules and three sulfate oxygen atoms, resulting in a total coordination number (CN) of ~7.95 (**Figure 5B**). The Yb–O bond lengths vary with CN, while mono-dentate sulfate oxygen forming the shortest bond (~2.38 Å) and bi-dentate sulfate oxygen and water oxygen exhibit longer distances (as shown the broader peaks at ~2.65 Å and ~3.03 Å in **Figure 5A**). These two asymmetric binding modes yield a distorted square antiprism geometry due to the uneven Coulombic repulsion between ligands and angular strain from bi-dentate sulfate ligands (as illustrated in **Figure 6A-B, E-F**). Although the beading strain for the molecular structures could destabilize the complex, the presence of π-bonding signature could compensate for these distortions. As shown in molecular orbital visualizations in **Figure 6C-D, G-H**, the electron density is always in a shoulder-to-shoulder shape, which suggests a synergistic overlap of ligand-based π orbitals with Yb, which could introduce extra covalency to the bonding and stabilize the molecule to exist in geometrically distorted forms without significant energetic penalties.

As the temperature increases to 100°C, the Yb–O RDF peaks broaden slightly, with minor elongation in bond lengths (sharp peak at ~2.39 Å for mono-dentate sulfate oxygen and broad peaks at ~2.67 Å and ~3.03 Å for bi-dentate sulfate and water oxygens). Despite these changes, the CN remains stable at ~7.90, indicating that the hydration shell is still intact at this temperature. The coordination geometry remains distorted, as the bi-dentate sulfate binding mode persists. The Yb–S RDF (**Figure 5C**) at 100°C shows that mono-dentate sulfate coordination (3.04 Å) becomes more prominent, while bi-dentate sulfate coordination exhibits a slight shift (~3.37 Å), suggesting increased thermal motion and dynamic exchange between sulfate and water ligands. Significant structural reorganization occurs at 200°C due to dehydration and increased thermal effects. The Yb–O CN decreases to ~7.19 (**Figure 5B**), reflecting the loss of one water molecule from the hydration shell. The Yb–O RDF (**Figure 5A**) shows broader and more variable bond lengths

(ranging from 2.26 Å to 3.13 Å), indicative of greater structural disorder. However, the Yb–S RDF (**Figure 5C**) demonstrates a pronounced peak at ~3.38 Å for bi-dentate sulfate coordination, suggesting stronger and more rigid sulfate binding. The decrease in Yb–S CN to ~1.54 (**Figure 5D**) highlights the reduced number of sulfate ligands directly coordinating with $Yb^{3+}$, yet less loss than the dehydration loss.

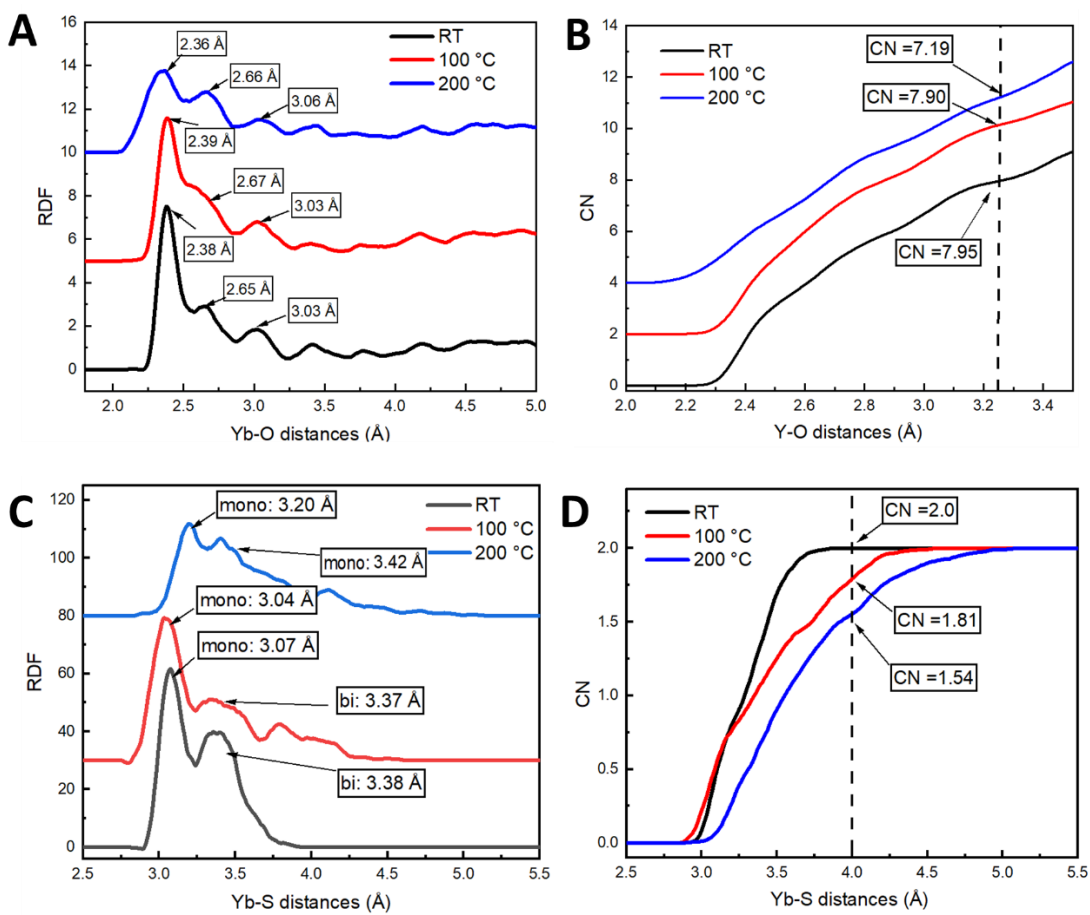

**Figure 5.** Illustration of the RDF (A and C) and total CN (B and D) for Yb-SO₄ at different temperatures: RT, 100 °C and 200 °C. In the Yb-S RDF illustration, the "mono" peak represents the $SO_4^{2-}$ ligand who binds with $Yb^{3+}$ via the monodentate mode, and the "bi" peak represents the $SO_4^{2-}$ ligand who binds with $Yb^{3+}$ via the bidentate mode. The Yb-S distance from the monodentate mode is always shorter than the Yb-S distance in the bidentate mode.

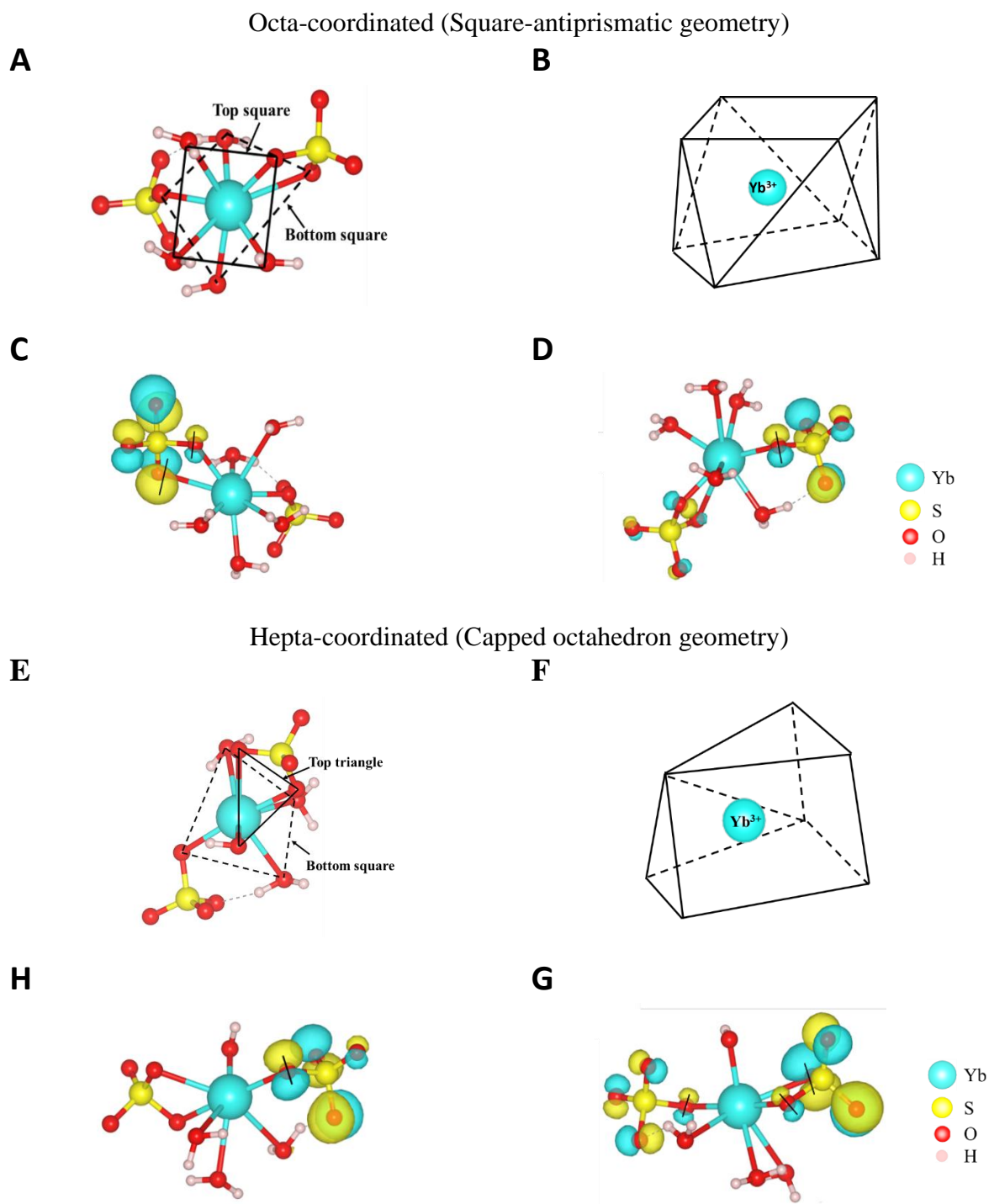

**Figure 6.** Display of octa- and hepta-coordinated Yb(III)-SO$_4$ complex from AIMD simulation. The snapshots are respectively from simulations at RT (octa-coordinated) and 200 °C. (A) Molecular structure of the octa-coordinated Yb complex with a square-antiprismatic geometry. (B) Polyhedral representation of the square-antiprismatic coordination geometry of Yb, showing the eight coordinating atoms forming two parallel squares. (C) Electron density visualization of a bonding orbital for the octa-coordinated Yb complex. (D) Electron density visualization of another bonding orbital for the octa-coordinated Yb complex. (E) Molecular structure of the hepta-coordinated Yb complex with a capped octahedron geometry. (F) Polyhedral representation of the capped octahedron coordination geometry of Yb. (G) Electron density visualization of a bonding orbital for the hepta-coordinated Yb complex. (H) Electron density visualization of another bonding orbital for the hepta-coordinated Yb complex. Cyan atoms represent Yb, yellow atoms represent S, red atoms represent O, and pale pink atoms represent H. Cyan electron density represent the electron accumulation, and the yellow electron density represent electron depletion.

## 4. Discussion

### 4.1 Geometry evolution of Yb$^{3+}$ sulfate complexes

The structural dynamics of Yb$^{3+}$-sulfate complexes show significant temperature-dependent evolution. At RT, Yb$^{3+}$ is coordinated by five water molecules and two sulfate ligands, resulting in a total CN of ~8.5. The coordination geometry corresponds to a distorted square antiprism, with one sulfate ligand binding in a bidentate mode and the other in a monodentate mode. This configuration produces a range of Yb–O bond lengths, with monodentate sulfate bonds at shorter distances (~2.36 Å) and bidentate bonds at longer distances (2.33–3.06 Å). The low Debye-Waller factors ($\sigma^2$) for Yb–O and Yb–S bonds indicate minimal thermal disorder, reflecting a stable and ordered coordination environment. These results are consistent across experimental and simulated datasets, underscoring the accuracy of AIMD-derived structural models. At 100°C, the coordination shell of Yb$^{3+}$ remains largely intact. Experimental data suggests slight elongation in Yb–O (~2.35 Å) and Yb–S (~3.68 Å) bond lengths, while the CN stable at ~8.5. However, the increase in $\sigma^2$ values for both Yb–O and Yb–S shells suggest enhanced thermal motion and dynamic ligand exchange. AIMD simulations further illustrate this dynamic behavior, showing increased flexibility in sulfate ligands and greater interaction with water molecules. Despite these fluctuations, the overall geometry remains stable, preserving the distorted square antiprism structure.

At 200°C, significant structural reorganization occurs due to thermal effects and partial dehydration. The experimental CN for Yb–O decreases to ~7.8, indicating the displacement of one water molecule from the hydration shell. AIMD simulations confirm this dehydration process and show that the remaining sulfate ligands bind more rigidly to $Yb^{3+}$. The Yb–S bond length remains stable (~3.64 Å), while the $\sigma^2$ for the sulfate shell decreases, suggesting reduced thermal motion and stronger interactions. This transition marks a shift from hydration-dominated coordination to sulfate-dominated coordination, driven by the chelating effect of bidentate sulfate ligands and reduced competition from water molecules.

At 300°C, the coordination environment evolves toward a structure resembling crystalline $Yb_2(SO_4)_3 \cdot 8H_2O$. Experimental EXAFS data show reduced Yb–O coordination (CN ~7.2) and increased Yb–S coordination (CN ~3.5), with shorter Yb–S bond lengths (~3.50 Å). These observations suggest a further decrease in hydration and a dominance of sulfate ligands in the coordination shell. This transition aligns with the retrograde solubility behavior of REE sulfates and hints at the onset of solid-phase formation.

### 4.2 Geological implications

The coordination environment of $Yb^{3+}$ observed in this study transitions from a hydration-dominated structure at RT to a sulfate-dominated structure at elevated temperatures, contrasting with previous studies on chloride ($Cl^-$) and fluoride ($F^-$) complexes. Chloride ligands exhibit weak electrostatic interactions with REEs, leading to more flexible ligand exchange and dynamic hydration shells. This results in a predominantly hydration-controlled transport mechanism, enabling REEs to remain mobile across a broad range of hydrothermal conditions (Mayanovic et al., 2007; Guan et al., 2020). In contrast, fluoride forms strongly bound inner-sphere complexes that restrict REE mobility and promote selective precipitation at relatively lower temperatures(Migdisov and Williams-Jones, 2008). Sulfate ligands, however, exhibit a dual behavior depending on temperature. At RT and 100°C, EXAFS and AIMD results confirm that sulfate coordinates with $Yb^{3+}$ in either a bidentate or monodentate mode, forming stable $Yb^{3+}$-$SO_4$ complexes in solution. At 200°C, the presence of two mono-dentate sulfate coordination observed suggests that sulfate strengthens its bond with $Yb^{3+}$ as the hydration shell diminishes—unlike chloride complexes, which tend to weaken at high temperatures. Above 200°C, sulfate may start to act as a deposition ligand, given that at 300°C the EXAFS data and AIMD simulations reveal a shift toward sulfate-dominated coordination, resembling the solid-phase $Yb_2(SO_4)_3 \cdot 8H_2O$ observed in diffraction studies (Hiltunen and Niinistö, 1976). The reduction in Yb-O coordination number (CN ~7.2) and increased Yb-S interaction (CN ~3.5) indicates the collapse of the hydration shell, favoring sulfate coordination over water. This suggests that sulfate facilitates REE precipitation under high-temperature conditions, consistent with the retrograde solubility of REE sulfates reported in natural hydrothermal deposits(Wan et al., 2023).

Unlike $Cl^-$, which remains a transport ligand across a wider P-T range, sulfate transitions from a stabilizing ligand in solution to a precipitating agent at high temperatures. This temperature-dependent transformation actively regulates REE solubility in hydrothermal fluids, ultimately influencing mineral deposition. The progressive dehydration observed from 100°C to 300°C enhances sulfate complexation strength, reducing REE solubility and triggering precipitation at higher temperatures. Geological evidence indicates that many REE deposits, such as Wicheeda (Canada) and Dalucao (China), formed due to abrupt T-P drops in sulfate-rich fluids (Trofanenko et al., 2016; Zhang et al., 2022). Fluid immiscibility further enhances REE concentration by partitioning REEs into sulfate-rich dense phases, a phenomenon observed in high-grade bastnäsite and monazite deposits (Wan et al., 2021). The increased rigidity of sulfate-bound REEs at high temperatures, as suggested by lower Debye-Waller factors ($\sigma^2$) in EXAFS fits, indicates that sulfate complexes play a stabilizing role at moderate temperatures but promote crystallization at higher temperatures. Overall, this temperature-dependent hydration-to-sulfate transition demonstrates that, in the absence of perturbations from other ions, sulfate can function both as a transport agent at lower temperatures and as a deposition trigger at higher temperatures—behavior distinct from chloride or fluoride systems.

**4.3 Geochemical Modeling**

The molecular-level insights gained from this study also have significant implications for geochemical modeling of REE behavior in hydrothermal systems. Accurate structural data on REE-sulfate complexes at high temperatures and pressures are essential for developing robust thermodynamic models that can predict REE speciation, solubility, and transport in natural hydrothermal fluids. These models are critical for understanding the formation of REE deposits and for guiding exploration and resource assessment strategies. For example, the retrograde solubility behavior of REE sulfates, as observed in this study, can be incorporated into thermodynamic databases to improve the accuracy of predictive models for REE transport and deposition. The retrograde solubility, where REE sulfate solubility decreases with increasing temperature, is a key factor in the precipitation of REE-rich minerals in hydrothermal systems. This behavior contrasts with the prograde solubility of many other metal complexes, highlighting the unique role of sulfate ligands in REE geochemistry.

The temperature-dependent transition from hydration-dominated to sulfate-dominated coordination provides a mechanistic basis for understanding the precipitation of REE-rich minerals in hydrothermal systems. At lower temperatures, sulfate ligands stabilize REEs in solution, facilitating their transport through hydrothermal fluids. However, as temperatures rise, the dehydration of REE complexes and the strengthening of sulfate binding led to the formation of more ordered, sulfate-rich solids. This transition is particularly relevant in high-temperature hydrothermal environments, where sulfate ligands can act as both transport and deposition agents. By integrating these insights into geochemical models, we can better predict the conditions under which REEs are likely to be transported or deposited, aiding in the identification of potential REE resources.

Moreover, the findings from this study can be used to refine existing thermodynamic models that simulate REE behavior in hydrothermal systems. For instance, the structural data on $Yb^{3+}$-sulfate complexes can be used to parameterize models that account for the effects of temperature, pressure, and fluid composition on REE speciation. These models can then be applied to natural systems, such as carbonatite-related REE deposits or hydrothermal veins, to better understand the processes controlling REE enrichment and mineralization. Additionally, the role of fluid immiscibility in concentrating REEs into sulfate-rich dense liquid phases can be incorporated into models to explain the formation of high-grade REE deposits, such as bastnäsite and monazite.

The integration of molecular-level insights into geochemical models also has practical applications in resource exploration. By understanding the conditions under which REE sulfates are stable or prone to precipitation, exploration efforts can be targeted toward areas with specific temperature, pressure, and fluid chemistry conditions that favor REE enrichment. Furthermore, these models can inform strategies for the sustainable extraction and recovery of REEs from hydrothermal ore deposits, contributing to the development of more efficient and environmentally friendly mining practices.

In conclusion, the molecular-level understanding of Yb$^{3+}$-sulfate complexes provided by this study offers valuable insights for geochemical modeling of REE behavior in hydrothermal systems. By incorporating these findings into thermodynamic models, we can improve our ability to predict REE transport and deposition, ultimately enhancing our understanding of REE ore formation and guiding the exploration and exploitation of these critical resources.

## 5. Summary

This study provides a detailed molecular-level understanding of Yb$^{3+}$ coordination in sulfate-rich hydrothermal fluids under high-pressure and high-temperature conditions. By integrating in situ EXAFS spectroscopy with AIMD simulations, we have characterized the structural evolution of Yb$^{3+}$ complexes in hydrothermal fluid up to 300°C. The results indicate a temperature-dependent transition in the Yb coordination environment, with hydration shell depletion and increasing sulfate interaction playing key roles in structural reorganization. At room temperature, Yb$^{3+}$ is coordinated by five water molecules and two sulfate ligands, forming a distorted square antiprism geometry with a coordination number of ~8. As the temperature rises, progressive dehydration occurs, reducing water coordination and strengthening sulfate binding. By 200°C, Yb$^{3+}$ adopts a capped dodecahedron configuration with a coordination number of ~7, and sulfate ligands become the dominant coordinating species. At 300°C, further dehydration and precipitation formed due to retrograde solubility of Yb$_2$(SO$_4$)$_3$, which also enhanced sulfate binding. These findings have significant geological implications in understanding REE transport and deposition in hydrothermal environments, which reveals sulfate acts as a critical ligand in REE mobility, stabilizing Yb$^{3+}$ in solution at lower temperatures while facilitating precipitation at higher temperatures due to stronger sulfate interactions. The transition from hydration-dominated to sulfate-dominated coordination provides new insights into the mechanisms driving REE mineralization, which could aid in refining predictive geochemical models for REE ore formation under extreme conditions.

## CRediT authorship contribution statement

**Xiaodong Zhao:** Experimental data curation, Calculation, Formal analysis, Investigation, Methodology, Writing – original draft, Writing – review & editing. **Duo Song**: Calculation, Methodology, Resources, Writing – review & editing. **Sebastian Mergelsberg:** Formal analysis, Investigation, Methodology, Writing – review & editing. **Micah Prange**: Formal analysis, Investigation, Methodology, Writing – review & editing. **Daria Boglaienko:** Experimental data curation, Methodology, Writing – review & editing. **Zihua Zhu**: Formal analysis, Writing – review & editing. **Zheming Wang**: Formal analysis, Writing – review & editing. **Carolyn I Pearce**: Methodology, Resources, Supervision, Writing – review & editing. **Chengjun Sun**: Methodology, Resources, Writing – review & editing. **Kevin M. Rosso**: Writing – review & editing, Supervision, Resources, Funding acquisition. **Xiaofeng Guo**: Conceptualization, Data curation, Formal analysis, Investigation, Methodology, Funding acquisition, Writing – review & editing. **Xin Zhang:** Conceptualization, Investigation, Methodology, Funding acquisition, Writing – review & editing.

## Declaration of competing interest

The authors declare that they have no known competing financial interests or personal relationships that could have appeared to influence the work reported in this paper.

## Acknowledgements

This work was supported by the U.S. Department of Energy (DOE), Office of Science, Basic Energy Sciences, Chemical Sciences, Geosciences, and Biosciences Division through its Geosciences Program at PNNL (FWP #67554). XZ and XZ also acknowledge support from a Laboratory Directed Research and Development Project at Pacific Northwest National Laboratory (PNNL) and the Advanced Research Projects Agency-Energy's (ARPA-E) Mining Innovations for Negative Emissions Resource Recovery (MINER) program with award number 0002707-1515, "Re‐Mining Red Mud Waste for $CO_2$ Capture and Storage and Critical Element Recovery (RMCCS‐CER)". CIP acknowledge support from IDREAM (Interfacial Dynamics in Radioactive Environments and Materials), an Energy Frontier Research Center funded by the U.S. Department of Energy (DOE), Office of Science, Basic Energy Sciences (FWP 68932). X.G. acknowledges the support of this work by the National Science Foundation (NSF), Division of Earth Sciences, under award No. 2149848. A portion of the work was performed with the user proposal 51382 using the Environmental and Molecular Sciences Laboratory (EMSL), a national scientific user facility at PNNL sponsored by the DOE's Office of Biological and Environmental Research. Simulations were performed using the National Energy Research Scientific Computing Center (NERSC) supported by the Office of Science of the U.S. DOE operating under Contract No. DE-AC02-05CH11231. PNNL is a multiprogram national laboratory operated by Battelle Memorial Institute under contract no. DE-AC05-76RL01830 for the DOE. This research used resources of the Advanced Photo Source (APS); a US Department of Energy (DOE) Office of Science user facility operated for the DOE Office of Science by Argonne National Laboratory under contract no. DE-AC02-06CH113577.

## Appendix A. Supplementary material

The following are the Supplementary material to this article: Evolution of total energy and temperature over simulation time at various temperatures from AIMD.